\documentclass[twocolumn,prb,showpacs,superscriptaddress,amsmath,amssymb,floatfix]{revtex4}
\usepackage[latin1]{inputenc}
\usepackage{graphicx}
\usepackage{dcolumn}
\usepackage{bm}

\begin{document}



\title{Ac magnetic susceptibility of a molecular magnet submonolayer directly patterned onto a micro superconducting quantum interference device}

\author{M. J. Mart\'{\i}nez-P\'{e}rez}
\affiliation{Instituto de Ciencia de Materiales de Arag\'on (ICMA), CSIC-Universidad de Zaragoza, Pedro Cerbuna 12, E-50009 Zaragoza, Spain}
\affiliation{Dpto. de F\'{\i}sica de la Materia Condensada, Universidad de Zaragoza, Pedro Cerbuna 12, E-50009 Zaragoza, Spain}

\author{E. Bellido}
\affiliation{Centro de Investigaci\'on en Nanociencia y Nanotecnolog\'{\i}a (CIN2, CSIC-ICN)
Edificio CM7, Esfera UAB, Campus UAB, E-08193 Cerdanyola del Vall\'es, Spain}

\author{R. de Miguel}
\affiliation{Laboratorio de Microscop\'{\i}as Avanzadas (LMA) - Instituto de Nanociencia
de Arag\'on (INA), Universidad de Zaragoza, 50018 Zaragoza, Spain}

\author{J. Ses\'e}
\affiliation{Laboratorio de Microscop\'{\i}as Avanzadas (LMA) - Instituto de Nanociencia
de Arag\'on (INA), Universidad de Zaragoza, 50018 Zaragoza, Spain}
\affiliation{Dpto. de F\'{\i}sica de la Materia Condensada, Universidad de Zaragoza, Pedro Cerbuna 12, E-50009 Zaragoza, Spain}

\author{A. Lostao}
\affiliation{Laboratorio de Microscop\'{\i}as Avanzadas (LMA) - Instituto de Nanociencia
de Arag\'on (INA), Universidad de Zaragoza, 50018 Zaragoza, Spain}
\affiliation{Fundaci\'on ARAID, Arag\'on, Spain}

\author{C. G\'omez-Moreno}
\affiliation{Laboratorio de Microscop\'{\i}as Avanzadas (LMA) - Instituto de Nanociencia
de Arag\'on (INA), Universidad de Zaragoza, 50018 Zaragoza, Spain}
\affiliation{Dpto. de Bioquímica, Universidad de Zaragoza, Pedro Cerbuna 12, E-50009 Zaragoza, Spain}

\author{D. Drung}
\affiliation{Physikalisch-Technische Bundesanstalt (PTB) Abbestraße 2-12, D-10587 Berlin, Germany}

\author{T. Schurig}
\affiliation{Physikalisch-Technische Bundesanstalt (PTB) Abbestraße 2-12, D-10587 Berlin, Germany}

\author{D. Ruiz-Molina}
\email{druiz@cin2.es} \affiliation{Centro de Investigaci\'on en Nanociencia y Nanotecnolog\'{\i}a (CIN2, CSIC-ICN)
Edificio CM7, Esfera UAB, Campus UAB, E-08193 Cerdanyola del Vall\'es, Spain}

\author{F. Luis}
\email{fluis@unizar.es} \affiliation{Instituto de Ciencia de Materiales de Arag\'on (ICMA), CSIC-Universidad de Zaragoza, Pedro Cerbuna 12, E-50009 Zaragoza, Spain}
\affiliation{Dpto. de F\'{\i}sica de la Materia Condensada, Universidad de Zaragoza, Pedro Cerbuna 12, E-50009 Zaragoza, Spain}


\date{\today}


\begin{abstract}

We report the controlled integration, via Dip Pen Nanolithography, of monolayer dots of ferritin-based CoO nanoparticles ($12 \mu_{\rm B}$) into the most sensitive areas of a microSQUID sensor. The nearly optimum flux coupling between these nanomagnets and the microSQUID improves the achievable sensitivity by a factor $10^{2}$, enabling us to measure the linear susceptibility of the molecular array down to very low temperatures ($13$ mK). This method opens the possibility of applying ac susceptibility experiments to characterize two-dimensional arrays of single molecule magnets within a wide range of temperatures and frequencies.

\end{abstract}


\maketitle



The ac magnetic susceptibility of magnetic nanoparticles and single molecule magnets (SMMs) provides useful information on their spin and magnetic anisotropy,\cite{Luis06} as well as on the magnetic relaxation mechanisms.\cite{Hernandez96,Chudnovsky98,Gatteschi06} Miniaturized superconducting quantum interference devices\cite{Awschalom88,Wernsdorfer95,Lam03,Cleziou06,Hao11} (SQUIDs) should eventually become capable\cite{Cleziou06,Bouchiat09} of measuring the magnetization reversal of a SMM ($\mu_{\rm i} \sim 20 \mu_{\rm B}$ for the archetypal Mn$_{12}$ molecule). However, detecting the linear response sets even more stringent conditions: at $T = 1$ K, a magnetic field $H = 24$ A/m ($0.3$ Oe) induces a magnetic polarization $\langle \mu \rangle \simeq 0.007 \mu_{\rm B}$ on the same Mn$_{12}$ cluster. Measuring the susceptibility of even a molecular monolayer represents therefore a considerable challenge, which requires to take the sensitivity of magnetic susceptometry beyond its actual limits.\cite{Awschalom88} To maximize the magnetic coupling between SMMs and the SQUID, molecular nanomagnets need to be deposited onto specific areas of the sensor.\cite{Cleziou06,Bouchiat09} Even though diverse techniques have been developed for structuring molecules and nanoparticles on sensors,\cite{Hao11,Lam08,Bogani09} such a controlled integration remains extremely challenging.

In the present work, we apply dip pen nanolithography (DPN) \cite{Piner99} to deposit monolayer dots of ferritin-based nanomagnets on the most sensitive areas of a microSQUID ac susceptometer. With its direct write capabilities, DPN is an attractive tool for the nanostructuration on surfaces and for controlling the number of units deposited.\cite{Rozhok07,Salaita07,Basnar09,Bellido10AM} The sample consisted of cobalt oxide nanoparticles, $\simeq 2$ nm in diameter, whose magnetic moment $\simeq 12 \mu_{\rm B}$ is close to that of typical SMMs.\cite{Gatteschi06} These particles (CoO@Apoferritin) are synthesized inside the protein nanocavity of horse spleen apoferritin\cite{Uchida10} and can be patterned and immobilized over different substrates.\cite{Bellido10scanning} The bulk magnetic susceptibility of this material was characterized using $\sim 10^{-9}$ Kg of CoO@Apoferritin. Further details of this and other experimental aspects are given in the supplementary material (see ref. \onlinecite{Supplementary}).

\begin{figure}[t]
\resizebox{7.5 cm}{!}{\includegraphics{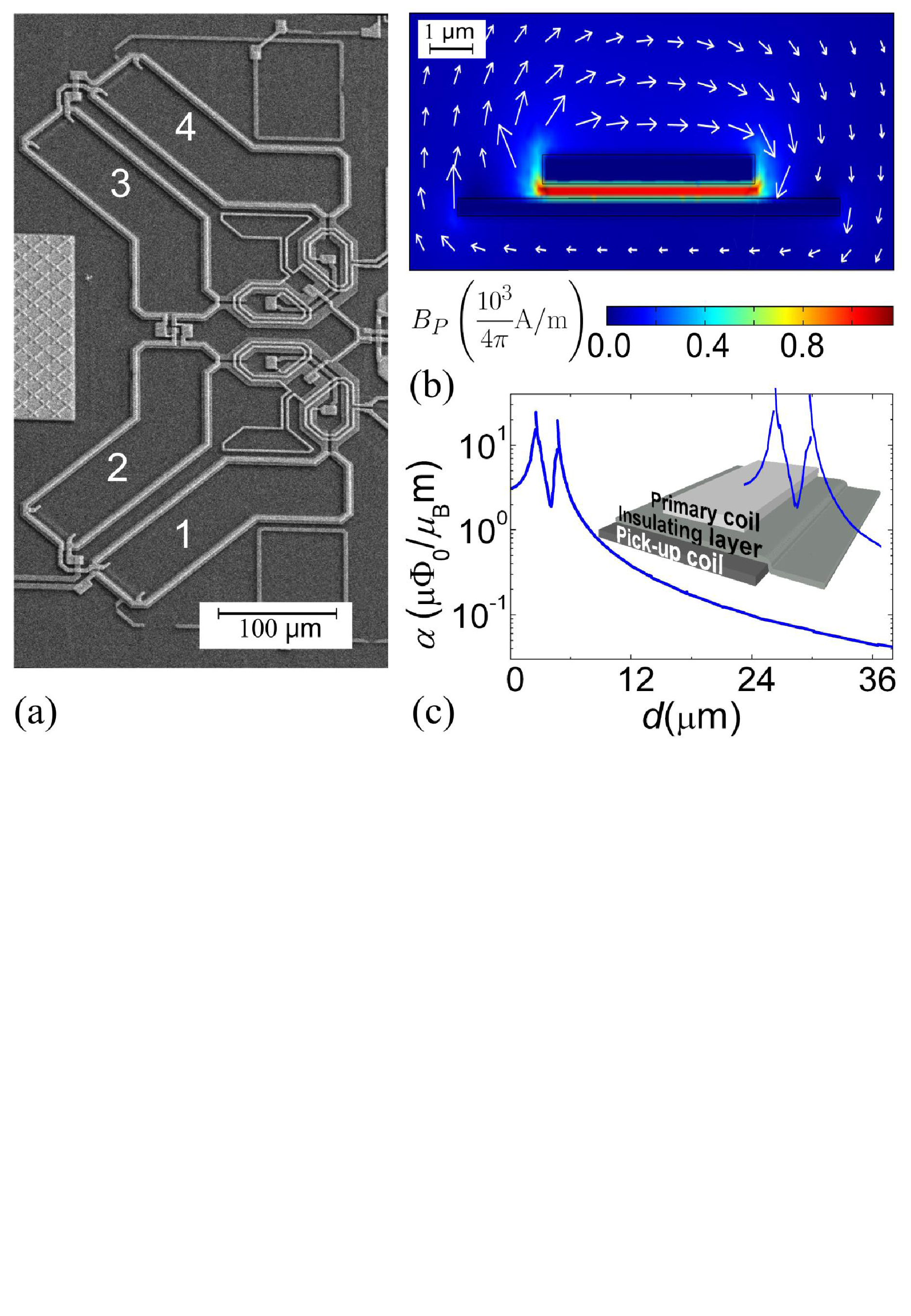}}
\caption{(a) SEM image of the SQUID showing the four rectangular shaped pick-up coils with effective areas of $63$ $\mu$m $\times$ $250$ $\mu$m. (b) Finite element calculation of the excitation magnetic field ($B_{\rm p}$) created by a $i_{\rm p} = 500$ $\mu$A current flowing through the primary coil, approximated by a circular spire. (c) Numerical calculations of $\alpha$ as a function of the distance from the center of the pick-up coil wire towards the center of the coil, approximated also by a circle. The inset shows a 3-D cross section of the pick-up and primary coil wires, where the $\alpha$ profile has been superimposed.}
\label{SQUID}
\end{figure}

The microSQUID susceptometer used for these studies has been described elsewhere.\cite{Supplementary,MartinezPerez10,MartinezPerez11} The pick-up coil most sensitive ("active") areas were identified by calculating (see Fig. \ref{SQUID}c and Ref. \onlinecite{Supplementary}) the magnetic flux $\Phi_{\rm coupled}$ generated by a sample located at a particular position. The coupling can be quantified by a flux coupling factor
\begin{equation}
\alpha = \frac{\Phi_{\rm coupled}}{\mu_{\rm i}}\frac{B_{\rm p}}{i_{\rm p}}
\label{alpha}
\end{equation}
\noindent where  $\mu_{\rm i}$ is the magnetic moment induced by the excitation magnetic field $B_{\rm p}$, and $i_{\rm p}$ is the electrical current circulating via the primary coil. We find that $\alpha$ can be enhanced by more than three orders of magnitude by simply placing the nanomagnets sufficiently close to the coil wire edges, where the magnetic field lines concentrate.

\begin{figure}[t]
\resizebox{7.5 cm}{!}{\includegraphics{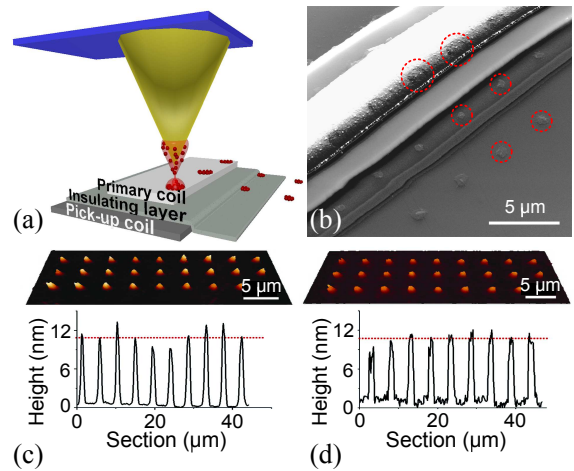}}
\caption{(a) Schematic representation of the nanoparticle deposition, by DPN, on the most active areas of the sensor. (b) SEM images of a sensor right after depositing three rows of CoO@Apoferritin dots. (c) and (d) AFM images and topographic profiles of CoO@Apoferritin dots deposited onto SiO$_{2}$ and Nb substrates, respectively.}
\label{DPN}
\end{figure}

The rational deposition of CoO@Apoferritin on these active areas is depicted in Fig. \ref{DPN}. Three rows of CoO@Apoferritin dots separated by $4$ $\mu$m were fabricated on the pick-up coils labeled $3$ and $4$ in Fig. \ref{SQUID}a by traversing the tip soaked with the ferritin-based nanoparticles over the specific areas. The first row was deposited on the primary Nb coil and the other two were deposited on the SiO$_{2}$ layer. The SEM images (Fig. \ref{DPN}b) reveal the high precision achieved in positioning the dots at the positions of maximum $\alpha$. The dots dimensions were measured by AFM (see Figs. \ref{DPN}c and d) on arrays deposited on bare SiO$_{2}$ and Nb substrates under identical conditions. We find average diameters of $1.3 \pm 0.1$ $\mu$m and $1.8 \pm 0.1$ $\mu$m for SiO$_{2}$ and Nb substrates, respectively. The average dot height was $11 \pm1$ nm in both, close to the size of a single protein (ca. $12$ nm), thus showing that each dot is a monolayer. According to these values, the average number of CoO@Apoferritin units per dot is $10^{4}$ and $2 \times 10^{4}$ for SiO$_{2}$ and Nb, respectively. The number of CoO@Apoferritin units deposited over the pick-up coils is $n \sim 10^{7}$.

\begin{figure}[h]
\resizebox{7 cm}{!}{\includegraphics{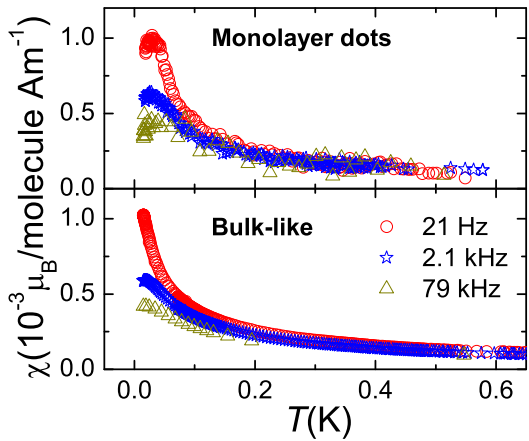}}
\caption{Top: In-phase ac magnetic susceptibility of $10^{7}$ CoO@Apoferritin molecules arranged as a (sub)monolayer. The out-of-phase component lies below the sensitivity limits of detection and it is therefore not shown. Bottom: In-phase susceptibility of $\sim 10^{-9}$ Kg of CoO@Apoferritin ($\sim 10^{12}$ units).}
\label{susc}
\end{figure}

The large coupling between the CoO@Apoferritin dots and the SQUID enabled us to measure their magnetic susceptibility down to $T = 13$ mK (Fig. \ref{susc}). Below $400$ mK, a temperature dependent signal shows up above the background signal of the bare sensor that was previously characterized.\cite{MartinezPerez11} Furthermore this signal shows the same qualitative dependence on temperature as the susceptibility $\chi^{\prime}$ of a bulk-like sample of CoO@Apoferritin measured with the same sensor under the same conditions.\cite{Supplementary} The magnetic polarization of the array can be estimated as $\langle \mu \rangle = n \chi^{\prime} B_{\rm p}$. Its maximum value, at $T \simeq 50$ mK, amounts to only $2.3 \times 10^{5} \mu_{\rm B}$.

Below approximately $100$ mK, $\chi^{\prime}$ depends on frequency. This shows the existence of a thermally activated spin reversal with characteristic timescale $\tau = \tau_{0} \exp(U/k_{\rm B} T)$, where $\tau_{0}$ is an attempt time and $U$ is the activation energy of the reversal process.\cite{Neel49,Brown63} When $\tau$ becomes comparable to $1/\omega$, the spins cannot follow in phase the oscillations of the excitation magnetic field. The maximum of $\chi^{\prime}$ {\em vs} $T$ that we observe for the array (see Fig. \ref{susc}a) defines the "blocking" temperature, characteristic of a SMM, which occurs when $\tau \gtrsim 1/\omega$.\cite{Hernandez96,Gatteschi06} Curiously enough, the bulk $\chi^{\prime}$ shows no clear maxima above $13$ mK. At first, this might suggest that the blocking temperature, thus also $U$, is enhanced in the array by the interaction of the molecules with the substrate. Alternatively, the temperature shift can be ascribed to a different thermalization of both samples. In the array, with its larger contact-area to volume ratio, the molecules can properly thermalize with the surrounding He bath. In contrast, the actual temperature of the bulk sample can stay above that of the He bath (and thermometer), therefore not reaching the blocking temperature. This interpretation is supported by the fact that the bulk susceptibility is shifted with respect to that of the array already at $T \lesssim 200$ mK, when $\chi^{\prime}$ is nearly independent of frequency and therefore relaxation mechanisms should not influence its temperature dependence.\cite{Supplementary}

Using these data it is possible to determine the average coupling factor $\alpha$. For this, we replace in Eq. (\ref{alpha}) $\mu_{\rm i}$ by the net polarization of the molecular array $\langle \mu \rangle$, defined above. The experimental $\Phi_{\rm coupled}$ can be determined from the SQUID's output voltage, since they are related trough fabrication parameters. Inserting real values in Eq. (\ref{alpha}) gives $\alpha = 28.6 (\pm 0.1) \mu\Phi_{0}/\mu_{\rm B}$m, of the same order of magnitude, albeit more than three times larger, as the average $\alpha = 8.0 (\pm 0.1) \mu\Phi_{0}/\mu_{\rm B}$m extracted from the numerical calculations shown in Fig. \ref{SQUID}.\cite{Supplementary} The discrepancy can be ascribed to the approximations made to simplify these calculations, in particular the use of circular primary and pick-up coils. This parameter gives a spin sensitivity $\simeq 300 \mu_{\rm B}$ Hz$^{-1/2}$ at $13$ mK, which represents an enhancement of two orders of magnitude with respect to the previous calibration performed with a $45$ $\mu$m thick Pb sphere.\cite{MartinezPerez11}

Summarizing, we have fabricated submonolayer arrays of ferritin-based nanomagnets ($12 \mu_{\rm B}$) on those regions that have a maximum flux coupling with a microSQUID loop. This controlled integration enhances the sensitivity by a factor $10^{2}$. Furthermore, the molecular deposition is carried out under ambient temperature and pressure conditions and implies no chemical functionalization of the sensor neither of the sample. The enhanced sensitivity has enabled us to directly measure the linear susceptibility of the molecular array, which shows that each molecule preserves its magnetic properties. The present technology opens the possibility of using the ac susceptibility to characterize two-dimensional arrays of single-molecule magnets. The same approach can be also applied to optimize the flux coupling of magnetic molecules to any other superconducting circuit, such as planar resonators, therefore contributing to the realization of hybrid architectures for quantum computation.\cite{Imamoglu09}


\begin{acknowledgments}
This work was partly funded by grant MAT2009-13977-C03 of the Spanish MICINN and the Consolider-Ingenio project on Molecular Nanoscience. M. J. M.-P. thanks CSIC for a JAE predoctoral fellowship. E. B. and R. d M. thank the MICINN for FPI and FPU predoctoral grants. A. L. thanks ARAID for financial support. We acknowledge helpful discussions with C. Carbonera and D. Maspoch. M. J. M. P. and E. B. contributed equally to the present work
\end{acknowledgments}



\end{document}